\documentclass{aa}
\usepackage{graphicx}
\begin{document}

%
\title{A wide field survey at the Northern Ecliptic Pole\\
II. Number counts and galaxy colours in $B_j$, $R$, and $K$
\thanks{Based on observations collected at the German-Spanish Astronomical
Centre, Calar Alto, operated by the Max-Planck-Institut f\"ur Astronomie,
Heidelberg, jointly with the Spanish National Commission for Astronomy}
\thanks{In partial fulfillment of the requirements for a Ph.D., carried out
at Landessternwarte Heidelberg}}
\titlerunning{Number counts and colours in $B_j$, $R$, and $K$}
\author{M.\ W.\ K\"ummel\inst{1, }\inst{2}
\and
S.\ J.\ Wagner\inst{2, }\thanks{Email:swagner@lsw.uni-heidelberg.de}}
\authorrunning{M.~K\"ummel \and S.~Wagner}
\institute{
Max-Planck-Institut f\"ur Astronomie,
K\"onigstuhl 17, D-69117 Heidelberg, Germany
\and
Landessternwarte Heidelberg-K\"onigstuhl, K\"onigstuhl 12,
D-69117 Heidelberg, Germany}
\offprints{M.W.\ K\"ummel}
\mail{kuemmel@mpia-hd.mpg.de}
\date{Received 22 May 2000/ Accepted 22 January 2001}
\abstract{
We present a medium deep survey carried out in the three filters $B_j$, $R$
and $K$. The survey covers homogeneously the central square degree around
the Northern Ecliptic Pole (NEP) down to a completeness limit of
$24.25$, $23.0$ and $17.5\,\mathrm{mag}$ in  $B_j$, $R$ and $K$, respectively.
While the near infrared data have been presented in the first paper of this
series, here we concentrate on the optical data and the results based on the
combined $B_jRK$-data. The unique combination of area and depth in our survey
allows to perform a variety of investigations based on homogeneous material
covering more than ten magnitudes in apparent brightness. We analyze the
number counts for point-like and extended sources in $B_j$ and $R$
to determine the slopes
in $\mathrm{d}\log N/\mathrm{d}m$ and to test for possible breaks therein.
While we can confirm the slopes found in previous works with a higher
statistical significance, the largest uncertainty remaining for the amplitudes
is galactic extinction. We determine the colour distributions of galaxies
in $B_j-R$ and $R-K$ down to $B_j=24.0$ and $K=18.0\,\mathrm{mag}$,
respectively. The distributions in both colours are modeled using galaxy
spectral evolution synthesis. We demonstrate that the standard
models of galaxy evolution are unable to reproduce the steady reddening trend in
$R-K$ despite flawless fits to the colour distributions in the optical
($B_j-R$). The $B_jRK$ data collected over a large area provides the
opportunity to select rare objects like candidates for high-redshift galaxies
and extremely red objects (EROs, $R-K>5.0$) and to determine their
surface density. Our EROs are selected at an intermediate magnitude range and
contain contribution from both galactic as well as extragalactic sources.
At $K<16.5\,\mathrm{mag}$, where a morphological classification is possible,
the stellar component dominates the sample.
}
\maketitle
\section{Introduction}
Understanding galaxy evolution relies on large and homogeneous sets of
data. Inhomogeneities introduced by stacking surveys with different depth
to maintain comparable numbers of sources over a large dynamic range are
an important limitation in testing those models of galaxy evolution that
are  designed to reproduce the number counts in different filters. Such
models have been developed extensively over the last decade (see Koo 
\& Kron \cite{koo92}).
Models incorporating luminosity evolution have been found to explain
the number counts down to the faintest levels (Metcalfe et al.~\cite{metcalfe},
Pozzetti et al.~\cite{pozzetti}). As pointed out by Gardner (\cite{gardner98}) 
however, the colour distribution contains more information about the state of 
evolution than the pure number counts, and the modeling of the colour 
distributions is a good test for the evolution models which explain number 
counts in individual filters. Up to now such modeling of galaxy colours 
has mostly been done for bright, nearby samples (Bertin \& Dennefeld 
\cite{bertin97}) or for deep samples with small number statistics
(Pozzetti et al.~\cite{pozzetti}, McCracken et al.~\cite{mccrack}).\\
We carried out medium-deep surveys in the optical and near-infrared
regimes. They cover one square degree and were
performed in the optical $B_j$- and $R$-bands as well as with a
near-infrared $K$ filter. The 95\% completeness limits in $B_j$, $R$,
and $K$ are $24.25\,\mathrm{mag}$, $23.0\,\mathrm{mag}$, and
$17.5\,\mathrm{mag}$, respectively. The near-infrared survey and
results based exclusively on $K$-data have been presented in K\"ummel
\& Wagner (\cite{kuemmel}) (hereafter paper I).
\begin{table}
\caption{The coordinates in the ecliptic, galactic and 
supergalactic systems and measures of ISM column density towards the NEP}
\label{tab:coo}
\begin{tabular}{llcll}
\hline
\noalign{\smallskip}
$\alpha_{2000}$:&$18$h$\:00$m$\:00.0$s&&$\delta_{2000}$:&$66^{\circ}\:33'\:38.6''$\\
$l_{II}$:&$96^{\circ}.38$&&$b_{II}$:&$29^{\circ}.81$ \\
$ SGL$:&$33^{\circ}.30$&&$SGB$:&$38^{\circ}.34$ \\
\multicolumn{2}{r} 
{HI-column-density\footnotemark}&$=$&\multicolumn{2}{l}{$3.9-4.7\times 
10^{20}\,$cm$^{-2}$}\\
\multicolumn{2}{r}
{IR-$100$-$\mu$-cirrus\footnotemark}&$=$&\multicolumn{2}{l}
{$1.8-3.0\,$MJy/steradian}\\
\multicolumn{2}{r}
{$E_{B-V}$\footnotemark}&$=$&\multicolumn{2}{l}{$0.05\,$mag }\\
\noalign{\smallskip}
\hline
\end{tabular}
\end{table}
\addtocounter{footnote}{-2}\footnotetext{Dickey \& Lockman (\cite{dickey})}
\stepcounter{footnote}\footnotetext{ISSA (\cite{issa})}
\stepcounter{footnote}\footnotetext{Schlegel et al.\ (\cite{schlegel})}
The complete coverage of the $K$-survey by the $R$-data, both in depth and 
in the large area, allows us to study the colour evolution
of our $K$-selected as well as optically selected sample with a high
statistical significance. We report on tests of the models proposed for
galaxy evolution, specifically, we study whether they can reproduce the 
galaxy colours and their variation with brightness.\\
One particular field of interest is to determine the surface densities
of extremely red objects (EROs) in an intermediate range of
magnitudes.  EROs are objects with $R-K>5.0\,\mathrm{mag}$ (Cimatti et
al.~\cite{cim}, Daddi et al.~\cite{daddi}) and include galactic and
extragalactic
populations. While the galactic EROs population are late type stars
(M6 or even later, see Leggett \cite{leggett} and Wolf et
al.~\cite{wolf}), there is no unique explanation for the extragalactic
EROs. Among the different scenarios (possibly all of which
{\bf contribute} on some not yet determined level) are galaxies with an
old stellar population at high redshift with a strong
$4000\,\mathrm{\AA}$ break. For a redshift of $z>0.85$ this break
falls between the $R$- and the $K$- filter bandpasses, resulting in
very red colours. Another suggestion for EROs are starburst galaxies
or active galactic nuclei at a redshift $1<z<2$. In this case reddening 
by interstellar dust alters the observed SEDs (Thompson et
al.~\cite{thomps}). A third possibility, which is probably less important
in the magnitude range covered in our survey are very distant 
quasars, where the Lyman break is redshifted to $\lambda>700\,\mathrm{nm}$. In 
all cases the objects lie at moderate to high redshifts
and give important clues on galaxy evolution and their star
formation history.
Evidence that the largest fraction of the extragalactic component of the
EROs-population are high-z ellipticals comes from the high clustering
amplitude suggested in recent surveys (Daddi et al.\ \cite{daddi}).\\
The EROs search in our medium deep and medium wide survey at the NEP
bridges the gap between the large area multi colour surveys like DENIS
(Epchtein et al.~\cite{epch}) or 2MASS (Skrutskie, et al.~\cite{skru})
and the deeper surveys on smaller fields, e.g. CADIS (Thompson et 
al.~\cite{thomps}, Huang et al.~\cite{huang}), or Daddi et al. (\cite{daddi}).
While we can continue the DENIS-search for low
mass stars to larger distances (Delfosse et al.~\cite{delfosse}), we
can detect the bright end of the extragalactic EROs population, which
are identified in deep surveys like CADIS.\\
The specific field used for our studies is the Northern Ecliptic Pole (NEP),
which is special in having been surveyed intensively by scanning satellites
(ROSAT, IRAS). Our deep counts shall be used to identify sources in deep 
X-ray and far-infrared surveys (Brinkmann et al.~\cite{brink},
Hacking \& Houck \cite{hack}) and study their broad band energy distribution.\\

\section{Observations and data analysis}
\subsection{Observations}
Two medium deep surveys were carried out with the 3.5-m telescope on
Calar Alto, Spain in two observing runs from July 21-25, 1993 and August 6-8, 
1994. During both campaigns the telescope was equipped with a TEK CCD (CA \#7)
in the prime focus. The CCD has $1024\times 1024$ pixels with an image scale
of $0.403''/\mathrm{pixel}$. The coordinates of the NEP and
other field parameters important for survey work are given in
Table \ref{tab:coo}. To cover the central square degree around the NEP
with the field of view (FOV) of $6.9\times 6.9\,\mathrm{arcmin}^2$, an equally
spaced grid of $9\times 9$ and $10\times 10$ exposures was taken in $B_j$ and
$R$, respectively. The exposure times of the individual frames are
$10\,\mathrm{min}$ in $B_j$ and $6\,\mathrm{min}$ in $R$. In order to
obtain homogeneous photometry over the whole field we carried out
a snapshot survey in both bands with short exposure times and a large
field of view in
photometric conditions. For this purpose we used the 2.2-m telescope on
Calar Alto with the focal reducer CAFOS (Meisenheimer \cite{meise}).
Together with the SITe CCD (CA \#1) CAFOS has a circular
FOV with $13\,\mathrm{arcmin}$ diameter and $0.531''/\mathrm{pixel}$.
In each band the snapshot survey was performed on a grid of $4\times 4$ 
exposures with integration times in $B_j$ and $R$ of  $3\,\mathrm{min}$ 
and $2\,\mathrm{min}$, respectively. Although the snapshot survey does not 
cover the complete
field of the deep exposures, there is sufficient overlap with {\it every}
frame of the deep survey to define a common photometric zeropoint.\\
All data were obtained using the so-called R\"oser-BV and R\"oser-R2 filters.
The R\"oser-BV filter ($\lambda_{center}=497.7\,\mathrm{nm}$ and
$\Delta \lambda = 155.9\,\mathrm{nm}$) is similar to the B$_j$ filter
(see Gullixson et al.\ \cite{gully}). The R\"oser-R2 (see R\"oser \&
Meisenheimer \cite{roeser}) avoids the strong OH emission lines at
$\lambda > 760\,\mathrm{nm}$, which contaminate the standard R
filters, with a sharp cutoff at $740\,\mathrm{nm}$.\\
In both surveys the median value of the full width
at half maximum (FWHM) of the point spread function (PSF) is $1.5''$.
\subsection{Standard reduction}
The single raw frames were de-biased and flat-fielded.
For every observing run a bias frame was constructed using frames with
an integration time of $0\,\mathrm{sec}$, taken with closed CCD-shutter.\\
A ``super-flat-field'' in each band 
was obtained by using all exposures taken in that filter. The de-biased
exposures were normalized and the super-flat-field
was computed from the  median of the data values in every pixel.\\
Dark-subtraction could be neglected since none of the CCDs displayed
significant dark-currents.
\subsection{Photometric calibration}
Photometric calibration was obtained by observing several standard fields
from Christian et al.\ (\cite{chris}) and Odewahn et
al.\ (\cite{odewahn}). The $B$-magnitudes of the stars listed there were
transformed to $B_j$ using the equations given by
Gullixson et al.\ (\cite{gully}).\\
We observed standard fields at different
airmasses in all nights when the snapshot survey was carried out.
Instead of computing a zeropoint for every exposure of the snapshot survey 
from the extinction curve individually, we used the large overlap between
adjacent frames to enhance the homogeneity of the photometry.
In every overlap region bright, unsaturated stars were identified in each
pair of neighboring exposures to determine the {\it differential
zeropoint} between the two exposures. Following the method developed by
Glazebrook et al.\ (\cite{glazebrook}) this system of differential
zeropoints for every overlap was then transformed to a single differential
zeropoint for each exposure. In photometric conditions those differential
zeropoints of the exposures originate from different extinction,
hence extinction correction is done explicitly.
We computed the {\it absolute zeropoint} of every exposure by adding
a constant value, which was determined from $\chi ^2$-minimization of
the differences between the zeropoints computed differentially and the
zeropoints derived from the extinction curve.\\
The zeropoints from the snapshot survey were then transferred to each field
of the deep survey individually, using several stars in each case.\\
While we did not find a significant colour term for the R\"oser-R2 filter,
the transformation
\begin{equation}
\label{equ:cterm}
B_j = b+0.23*(B_j-R)
\end{equation}
was applied to convert the instrumental magnitude $b$ to $B_j$.
\subsection{Object detection and photometry}
\label{chap:magn}
Object detection, photometry and morphological classification was carried out
with FOCAS (Valdes \cite{valdes94}, Jarvis \& Tyson \cite{jarvis}).
The reliability of the detection process was extensively tested on
simulated images generated with the iraf-package {\it noao.artdata}.
As in paper I, the threshold parameters were chosen such that
only $<1\%$ of the objects found in the artificial images were {\it not}
real objects, resulting in a reliability of $>99\%$ for the detected objects.
This was achieved by setting the FOCAS parameters such that after a
convolution with the FOCAS built-in digital filter, at least nine connected
pixels are required to have intensities $\ge 2.8\,\sigma$ to be recorded
as an object.\\
Similar to paper I we used the FOCAS-total flux $L_{total}$ and the
flux measured in an aperture $L_{lfca}$ for bright and faint sources,
respectively (see paper I for a detailed discussion).
The transition from $L_{total}$ to $L_{lfca}$ was chosen
at an object size $\le 28\,\mathrm{arcsec}^2$. For those objects the
flux $L_{lfca}$ was measured in the corresponding aperture of $6''$ diameter.
\subsection{Morphological classification}
\label{chap:class}
\begin{figure}
\resizebox{\hsize}{!}{\includegraphics*[angle=-90]{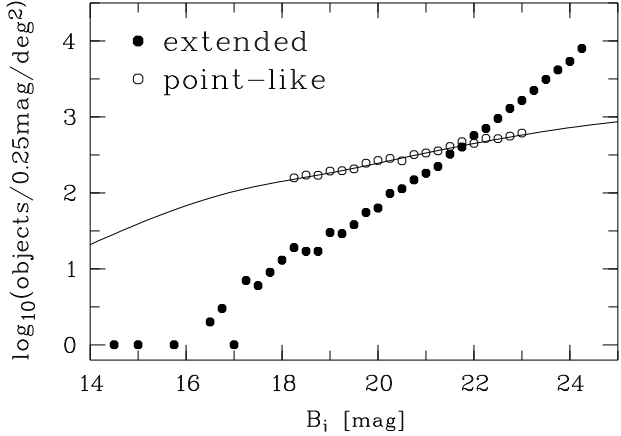}}
\caption{The number counts for extended and  point-like
sources in $B_j$. The solid line is the expected star counts
according to the Bahcall-Soneira model}
\label{fig:fig1}
\end{figure}

The morphological classification into point-like and extended sources
was done in both the $B_j$- and the $R$ survey. It is based on the
FOCAS-resolution classifier (Valdes \cite{valdes82}).
This classifier fits a series of templates, which are basically 
derived by scaling the width of the image PSF to each object. The scale
of the best fitting template is then a measure of the resolution of the
object and the classification is made from this scale value.\\
Like every classifier based on the object shape, the FOCAS-{\it resolution
classifier} does {\it not} give reliable results for sources near
the completeness limit. Towards lower signal-to-noise ratios the extended
parts of galaxies progressively vanish in the background noise.
This leads to a misclassification of extended sources as point-like objects.
This misclassification affects sources closer than $1.5\,\mathrm{mag}$ to
the completeness limit.\\
Point-like sources are stars, distant quasars, and nuclei of galaxies at low
and intermediate redshift which have steep luminosity profiles such that the 
width of the nuclear profiles down to the level of sky noise is significantly 
smaller than the PSF.
Down to our levels of completeness, the surface density of quasars is about
$85\,\mathrm{deg}^{-2}$, which is more than an order of magnitude lower
than the density of stars according to the Bahcall-Soneira
(Bahcall \& Soneira \cite{bahcall1}, Bahcall \cite{bahcall2},
Mamon \& Soneira \cite{mamon}) model of the Galaxy.
Nucleated dwarf galaxies, such as M32, would be included in our survey out
to distances of $200\,\mathrm{Mpc}$. The scale length of the nucleus would be 0.2", and
the surface brightness of the extended emission would be lost in the sky noise.
Such nucleated dwarf galaxies would be classified as point-like sources for
90 \% of the volume sampled. It is still unlikely, that dwarf galaxies present
a significant contribution to the point-like sources, since our survey only
includes one major galaxy out to the distance of 200 Mpc. The majority of 
point-like objects that are not stars are distant galaxies with small angular 
size. In the Hubble deep fields (HDF, HDFS) 12 \% of all extended objects down 
to $R=23\,\mathrm{mag}$ have scale-length that would render them
unresolved at our resolution. At brighter magnitudes this ratio cannot be
determined reliably due to small number statistics.\\
To derive the number counts of extended objects we used a statistical
source-classification in the range of unreliable FOCAS-classification.
We extrapolate the number-counts of bright point-like sources
by assuming $\mathrm{d}\log N/\mathrm{d}m=const$ and compute the number
of extended objects by subtracting the expected number of point-like sources
from the total number of objects. To test the assumption for the statistical
classification we computed the number of stars {\it expected}
according to the Bahcall-Soneira
model in $B_j$ and $R$.
Figs.\ \ref{fig:fig1} and \ref{fig:fig2} display the number
counts of point-like sources in $B_j$ and $R$ as open circles. Shown as a
solid line are the counts expected for the Bahcall-Soneira model.
There is a good agreement between the theoretically expected 
number of stars and the detected number of point-like sources in the range
of reliable classification. We actually
detect more point-like sources than expected, confirming a contribution of
up to ten percent of unresolved galaxies to the list of point-like objects.
For the stars no significant change in $\mathrm{d}\log N/\mathrm{d}m$
down to $B_j=24.25\,\mathrm{mag}$ and $R=23.0\,\mathrm{mag}$ is expected
in the model. This confirms the validity of the assumption in the
range of unreliable FOCAS-classification and justifies our
statistical classification. In the following we will treat point-sources
as stars, despite the above mentioned contamination.
\begin{table}
\caption{The source counts in $B_j$}
\label{tab:Bcoun}
\begin{tabular}{crrrrr}
\hline
\noalign{\smallskip}
Filter & mag& N$_{gal}$ & $\sigma _{N_{gal}}$& N$_{star}$& area\\
\noalign{\smallskip}
\hline
\noalign{\smallskip}
$B_j$ &   14.625 &1.00 &    1.00 & &        1.0\\
 &   15.125 &     1.00 &    1.00 & &        1.0\\
 &   15.625 &     1.00 &    1.00 & &        1.0\\
 &   16.125 &     0.00 &    0.00 & &        1.0\\
 &   16.625 &     5.02 &    2.25 & &        1.0\\
 &   17.125 &     8.03 &    2.83 & &        1.0\\
 &   17.625 &    15.07 &    3.89 & &        1.0\\
 &   18.125 &    32.15 &    5.67 & &        1.0\\
 &   18.625 &    34.16 &    5.84 &  342.58& 1.0\\
 &   19.125 &    59.27 &    7.70 &  389.80& 1.0\\
 &   19.625 &    93.43 &    9.66 &  453.09& 1.0\\
 &   20.125 &   161.74 &   12.72 &  551.54& 1.0\\
 &   20.625 &   262.21 &   16.19 &  583.69& 1.0\\
 &   21.125 &   404.87 &   20.12 &  693.20& 1.0\\
 &   21.625 &   727.36 &   26.97 &  879.05& 1.0\\
 &   22.125 &  1273.87 &   35.70 &  970.48& 1.0\\
 &   22.625 &  2250.38 &   47.44 & 1075.96& 1.0\\
 &   23.125 &  3885.19 &   62.33 & &        1.0\\
 &   23.625 &  7286.75 &   85.36 & &        1.0\\
 &   24.000 & 10775.98 &  146.80 & &        0.8\\
 &   24.250 & 15936.32 &  178.52 & &        0.3\\
\noalign{\smallskip}
\hline
\end{tabular}
\end{table}
\begin{table}
\caption{The source counts in $R$}
\label{tab:Rcoun}
\begin{tabular}{crrrrr}
\hline
\noalign{\smallskip}
Filter & mag& N$_{gal}$ & $\sigma _{N_{gal}}$& N$_{star}$& area\\
\noalign{\smallskip}
\hline
\noalign{\smallskip}
$R$ &   12.875& 0.97&    0.97 & &        1.0\\
 &   13.375&    0.00&    0.00 & &        1.0\\
 &   13.875&    0.97&    0.97 & &        1.0\\
 &   14.375&    0.00&    0.00 & &        1.0\\
 &   14.875&    1.94&    1.37 & &        1.0\\
 &   15.375&    3.88&    1.94 & &        1.0\\
 &   15.875&   11.65&    3.37 & &        1.0\\
 &   16.375&   22.33&    4.66 &  288.70& 1.0\\
 &   16.875&   37.88&    6.06 &  362.09& 1.0\\
 &   17.375&   45.64&    6.66 &  393.30& 1.0\\
 &   17.875&   62.15&    7.77 &  490.41& 1.0\\
 &   18.375&  132.07&   13.03 &  551.59& 1.0\\
 &   18.875&  239.86&   15.26 &  624.42& 1.0\\
 &   19.375&  390.39&   19.46 &  757.47& 1.0\\
 &   19.875&  652.59&   25.16 &  911.88& 1.0\\
 &   20.375& 1028.41&   31.60 &  960.43& 1.0\\
 &   20.875& 1691.10&   40.52 & 1199.05& 1.0\\
 &   21.375& 2460.88&   48.88 & 1386.00& 1.0\\
 &   21.875& 3420.84&   57.63 & &        1.0\\
 &   22.375& 5406.16&   72.45 & &        1.0\\
 &   22.875& 9252.67&   94.78 & &        0.6\\
\noalign{\smallskip}
\hline
\end{tabular}
\end{table}
\begin{table}
\caption{The source counts in $K$}
\label{tab:Kcoun}
\begin{tabular}{crrrrr}
\hline
\noalign{\smallskip}
Filter & mag& N$_{gal}$ & $\sigma _{N_{gal}}$& N$_{star}$& area\\
\noalign{\smallskip}
\hline
\noalign{\smallskip}
$K$ &    7.25 &     &        &   1.93 & 0.9\\
 &    7.75 &        &        &   1.93 & 0.9\\
 &    8.25 &        &        &   5.78 & 0.9\\
 &    8.75 &        &        &   5.78 & 0.9\\
 &    9.25 &        &        &   6.74 & 0.9\\
 &    9.75 &        &        &  10.59 & 0.9\\
 &   10.25 &        &        &  17.34 & 0.9\\
 &   10.75 &    0.92&   1.02 &  23.12 & 0.9\\
 &   11.25 &    0.00&   0.00 &  46.23 & 0.9\\
 &   11.75 &    1.93&   1.44 &  52.01 & 0.9\\
 &   12.25 &    2.89&   1.76 &  86.68 & 0.9\\
 &   12.75 &    2.89&   1.76 & 118.47 & 0.9\\
 &   13.25 &   15.41&   4.07 & 182.03 & 0.9\\
 &   13.75 &   20.23&   4.66 & 275.45 & 0.9\\
 &   14.25 &   34.67&   6.11 & 341.91 & 0.9\\
 &   14.75 &   86.68&   9.65 & 450.74 & 0.9\\
 &   15.25 &  112.69&  11.01 & 549.95 & 0.9\\
 &   15.75 &  265.82&  16.91 & 723.31 & 0.9\\
 &   16.25 &  488.31&  22.91 & 889.93 & 0.9\\
 &   16.75 &  911.44&  31.65 &1093.54 & 0.9\\
 &   17.25 & 1462.21&  48.96 &1214.90 & 0.6\\
\noalign{\smallskip}
\hline
\end{tabular}
\end{table}
\subsection{Completeness}
\label{chap:compl}
The completeness function $f$ is derived as described in paper I. We fit
\begin{equation}
f(mag) = \left[ \exp \left( \frac{mag-mag_{50}}{b}\right) +1 \right]^{-1}
\end{equation}
to the normalized number counts N$(m)$/\~N$(m)$ to determine $mag_{50}$, 
where the number of detected object is half the expected one (as inferred 
from the normalization). The parameter $b$ was found to be independent
within the range of image quality experienced in our survey at $b=0.26$,
while $mag_{50}$ varied in both data sets ($B_j$ and $R$), reflecting the
different image quality for the individual frames.
We established the relationship between $mag_{50}$ and the basic parameters 
of image quality, i.e.\ background and FWHM of the PSF for the $B_j$- and the
$R$-data independently, and used this relationship and the corresponding
zeropoint to calculate $mag_{50}$ for every survey image. Finally the
completeness limit for each image is set at $mag_{compl} = mag_{50}-0.6$.
At this level the completeness function is $f(mag_{compl})>0.9$, and the
slope-normalized number counts are indistinguishable from 1.0.\\
As in paper I the completeness function $f(mag)$ was not used to correct
number counts fainter than $mag_{compl}$. However objects down to
$mag_{50}$ were taken to match the objects found in different filters and to
determine the colours of objects. This is justified since those
sources in the range $[mag_{compl}-m_{50}]$ which have actually been found,
form a statistically selected subsample of sources detected with the high
reliability of $99\%$, even if not the entire population is included.
Down to $m_{50}$ of the individual fields we detect $58\,500$, $53\,900$ 
and $13\,000$ sources in the $B_j$-, $R$- and $K$-survey, respectively.
\subsection{Astrometry}
To obtain a precise absolute astrometry for the detected objects the
{\it Guide Star Catalog 1.2} (GSC) was used (Lasker et al.\ \cite{lasker},
Russel et al.\ \cite{russel}, Jenkner et al.\ \cite{jenkner}). There are 430
GSC-stars in the area covered by our surveys, and the number of GSC-stars
per exposures ranges from $1$ to $11$. All GSC-stars except the brightest
one (the planetary nebula NGC\,6543, $m_V=9.8\,\mathrm{mag}$) were taken to
establish the astrometry. We identified the GSC-stars on the survey images
and determined plate constants for every observing run and filter using
gnomonic projection (see Eichhorn\ \cite{eichhorn}). Using the plate constants
and the position of the GSC-stars we then computed the equatorial position
of every image center, and, finally, the equatorial positions of all objects
found on the images.\\
Because of the different spacing of the exposures
the position of an object in $B_j$ and $R$ is based on a different set
of GSC-stars. The astrometry can therefore considered to be independently
derived for the $B_j$- and the $R$-survey.
We took advantage of this in order to compensate for the
variable number of GSC-stars per image and to enhance the overall homogeneity
of the astrometry. For every overlap between an individual $B_j$-frame B$_k$
and an individual $R$-frame R$_l$ we computed $\Delta \alpha_{kl}$ and
$\Delta \delta_{kl}$, the mean offset between the $B_j$- and $R$-coordinates 
of bright stars in right ascension and declination, respectively.
Then we derived the corrections
$\Delta \alpha _{R_k}=\sum_l \Delta \alpha_{kl}$ and 
$\Delta \delta _{R_k}=\sum_l \Delta \delta_{kl}$
($\Delta \alpha _{B_l}=\sum_k (-\Delta \alpha_{kl})$ and 
$\Delta \delta _{B_l}=\sum_k (-\Delta \delta_{kl})$ for B$_l$) and applied them
to all positions in the whole frame. This procedure was iterated once to
minimize the contribution of a frame with bad astrometry to its overlaps.\\
We estimated the accuracy of the astrometry using bright stars in the large
($\sim 1'$) overlap regions between adjacent fields of the $R$-survey.
The accuracy of the object positions was, even for the faintest sources,
determined to be $<0.8''$ in each coordinate.
\begin{figure}
\resizebox{\hsize}{!}{\includegraphics*[angle=-90]{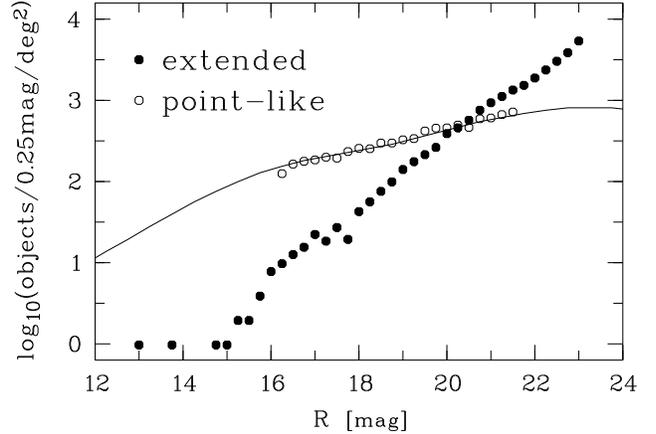}}
\caption{The number counts for extended and  point-like
sources in $R$. The solid line is the expected star counts
according to the Bahcall-Soneira model}
\label{fig:fig2}
\end{figure}
\subsection{Removal of cosmic ray objects}
Since most of the survey area is covered only by one frame, the usual approach
to remove cosmics with weighting maps or masks
(Nonino et al.\ \cite{nonino}, Arnouts et al.\ \cite{arnouts})
could not be followed. Instead we identified sources caused by cosmic
ray events and removed them from the object lists.\\
The criterion whether an object is real or just a cosmic ray hit
is the concentration of the brightness distribution
\begin{equation}
conc = 2.5*\log((Lc/9.0-ispht)/(ssbr/3.0))+mag
\end{equation}
The PSF sets an upper limit in the brightness concentration for real
objects. Cosmic ray hits usually have higher values of $conc$,
since their shape is {\it not} determined by the PSF and a cosmic ray
event usually affects only a few pixels.
$Conc$ is easily derived from the FOCAS-parameters $Lc$ (core luminosity),
$ispht$ (isophotal brightness) and $ssbr$ (sky-noise) together with
the object brightness $mag$ (the nomenclature of the FOCAS-parameters
follows Valdes \cite{valdes82}).\\
Looking at the $conc$-$mag$ distribution of all objects from an individual
image the locus of cosmic ray objects could be identified easily
and the objects could then be removed from the lists.
\begin{figure}
\resizebox{\hsize}{!}{\includegraphics*[angle=-90]{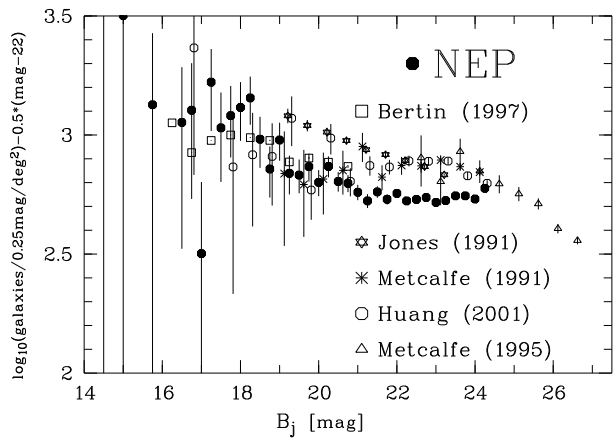}}
\caption{The galaxy counts in $B_j$ from the NEP compared to counts from
various other surveys}
\label{fig:fig7}
\end{figure}
\section{Object counts}
\label{chap:ncounts}
\subsection{Extended and point-like sources in $B_j$ and $R$}
Figs.\ \ref{fig:fig1} and \ref{fig:fig2} display the object counts in
$B_j$ and $R$, respectively. In both figures the counts of point-like
sources are marked with open circles, the counts for extended sources are denoted
with the filled symbols. We made no attempt to correct the counts beyond
the incompletentness of the individual frames. The counts were only derived
from fields complete to the specific magnitude.
Tables \ref{tab:Bcoun} and \ref{tab:Rcoun}
give the number counts (in $objects/0.5\mathrm{mag}/\mathrm{deg}^2$) $B_j$ and
$R$, respectively. The counts for extended objects are listed in column $3$ 
and counts for point-like objects in column $5$.
The fourth column reflects pure Poissonian error of the counts for extended 
objects. The last column gives the area of the sub-survey complete to
the specified depths. To complete the information concerning number
counts in our surveys we added the corresponding data for the $K$-survey
in Table \ref{tab:Kcoun}.\\
The point-like sources in Figs.~\ref{fig:fig1}, \ref{fig:fig2} and in
Tables \ref{tab:Bcoun}-\ref{tab:Kcoun} are only given down to the magnitude
of reliable
classification. The deeper counts of extended objects have been
derived by subtracting the expected number of point-like objects from
the counts of all objects (see Sect.~\ref{chap:class} and below).  No
number densities can be given at the bright end of the point-like
sources because the objects saturated the CCD. The solid line in
Figs.\ \ref{fig:fig1} and \ref{fig:fig2} shows the theoretically
expected stellar counts according to the Bahcall-Soneira model
(Bahcall \& Soneira \cite{bahcall1}, Bahcall \cite{bahcall2}). To
calculate the model counts in $B_j$ we transferred the original
$V$-counts with the model $B-V$-colours and equations given by
Gullixson et al.~(\cite{gully}). For the $R$-counts we changed the
code according to Mamon \& Soneira (\cite{mamon}).  No attempts were
made to improve the fit to our data by changing the parameters of the
model. This is justified by the good agreement between our counts and
the model.\\
The Bahcall-Soneira model does not take into account the so called
{\it thick disc} introduced by Gilmore \& Reid
(\cite{gilmore}). However, the scope of the comparison done in this
paper is not to test a particular model of the Galaxy. The agreement
between the counts for point-like objects and the Bahcall-Soneira
model is taken as confirmation of the statistical classification
applied to the total counts to derive the fraction of extended objects
(see chap.~\ref{chap:class}).
\subsection{Number densities of galaxies}
\begin{figure}
\resizebox{\hsize}{!}{\includegraphics*[angle=-90]{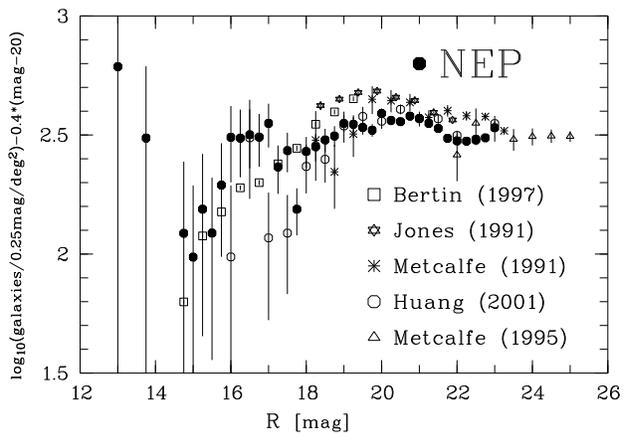}}
\caption{The galaxy counts in $R$ from the NEP compared to counts from
various other surveys}
\label{fig:fig8}
\end{figure}
In Figs.\ \ref{fig:fig7} and \ref{fig:fig8} we compare counts of galaxies
at the NEP with published counts from other surveys in $B_j$ and $R$, 
respectively.
The reference data are from Bertin \& Dennefeld (\cite{bertin97}),
Jones et al.\ (\cite{jones}), Huang et al.\ (\cite{huang}) and Metcalfe et
al.\ (\cite{metcalfe91}, \cite{metcalfe}). All $B_j$-counts except
Huang et al.\ (\cite{huang}) were either performed in a $B_j$-filter or transformed
to $B_j$ using equations given by the authors. To the Huang et
al.\ (\cite{huang}) {\bf photometry} we applied the transformation $B_j = B-0.19\,\mathrm{mag}$,
according to Bertin \& Dennefeld (\cite{bertin97}) for $B-R=1.0\,\mathrm{mag}$.
In the Figs.\ \ref{fig:fig7} and
\ref{fig:fig8} the slope $0.5 $ and $0.4$ is subtracted from the logarithm
of the counts to expand the ordinate and to make differences between the
counts clearly visible. In order to do a quantitative comparison we
fitted power-laws of the form
\begin{equation}
N(mag) = a*10^{b*(mag-c)}
\end{equation}
to the data. Tables \ref{tab:bjdurch} and \ref{tab:rdurch} show the results
of those fits. Both Tables give the literature reference and the survey area,
{\bf respectively}, (in deg$^2$)
in the first two columns. The next columns show the slope $b$, amplitude $a$
and the magnitude range in which the fits were done. In $B_j$ and $R$ the
fits were done for $c=22.0\,\mathrm{mag}$ and $20.0\,\mathrm{mag}$,
respectively. The faintest
magnitude bins are not included in the fit. The numbers in those bins might
be affected by {\it Eddington-bias} (Eddington \cite{eddi}),
and could influence the fits presented
in Tables \ref{tab:bjdurch} and \ref{tab:rdurch} significantly since their
high values are associated  with small (relative) errors.\\
We fitted different power-laws above and below $19.4\,\mathrm{mag}$ to
the $R$-band data from the NEP, since there is a clear break in the
number counts at this level.  While the slope is almost $0.5$ for the
bright magnitudes, it flattens by more than $0.1$ towards fainter
magnitudes.\\
The slopes in $B_j$ are in good agreement with other surveys with the
exception of Metcalfe et al. (\cite{metcalfe}). As can be seen in 
Fig.~\ref{fig:fig7}, the change in slope at $B\sim 24$ (Arnouts et
al.\ \cite{arnouts}, Williams et al.\ \cite{williams}) flattens the
slope in the deep surveys of Metcalfe et al.\ (\cite{metcalfe}).
While the slope of the NEP-counts in $B_j$ is comparable to the slope
of other surveys to the same limiting magnitudes, the amplitude is at
$a=566$ very low compared to the others, which show values around $720$.\\
In $R$ the NEP-counts clearly resolve the break in the slope at
$19.0\,\mathrm{mag}$ leading from $0.5$ at the bright counts to $0.37$ at the
faint end. As is the case in $B_j$ the slopes at the NEP agree well with
the other surveys, but the amplitudes are lower.\\
The main reason for the low amplitudes in both the $B_j$- and the $R$-filters
can be attributed to the low galactic latitude and therefore high extinction
value. While typical extragalactic survey fields used e.g. in
Metcalfe et al.\ (\cite{metcalfe}) have $E_{B-V}=0.02$ the extinction
at the NEP is $E_{B-V}=0.05$. This can be translated into a fading
of $0.12\,\mathrm{mag}$ and $0.07\,\mathrm{mag}$ (Schmidt-Kaler
\cite{schmika}) of the NEP-sources with
respect to sources from other surveys in $B_j$ and $R$, respectively.
This is supported by the fact that at the longest wavelength $K$
there is no such effect (see paper I) while the difference in the amplitude
is strongest in $B_j$, the shortest wavelength.\\
Assuming all of the offset is due to extinction, would lead us to shift
our $B_j$- and $R$-counts by $\Delta B_j=0.2$ and $\Delta R =
0.15\,\mathrm{mag}$, respectively, corresponding to an extinction
of $\Delta E_{B-V} = 0.05$ {\it more} than those adopted for 
the fields observed e.g.~ by Metcalfe et al.\ (\cite{metcalfe}).
The extinction at the NEP then would have to be $E_{B-V}=0.07$,
$0.02$ higher than the values from Schlegel et al.~(\cite{schlegel})
(see Table \ref{tab:coo}), or the extinction towards the Metcalfe et
al.~(\cite{metcalfe}) fields would have to be negligibly small. 
If the extinction given by Schlegel et al.~(\cite{schlegel}) is taken into account,
the difference between the counts at the NEP and other surveys reduces
to an acceptable amount of $\sim 10\%$.
\begin{table}
\caption{A quantitative comparison of galaxy counts in different $B_j$-surveys}
\label{tab:bjdurch}
\begin{tabular}{ccccc}
\hline
\noalign{\smallskip}
survey&area&slope&amplitude&range\\
\noalign{\smallskip}
\hline
\noalign{\smallskip}
Bertin&$140$&$0.464\pm 0.001$&$663\pm 5$&16.0-21.0\\
Jones&$2.1$&$0.442\pm 0.003$&$801\pm 5$&18.96-23.46\\
NEP&$1.0$&$0.479\pm 0.005$&$566\pm 7$&14.38-23.63\\
Metc.$_{91}$&$0.079$&$0.491\pm 0.009$&$748\pm 27$&18.88-24.38\\
Huang&$0.19$&$0.473\pm 0.006$&$764\pm 18$&16.75-24.75\\
Metc.$_{95}$&$0.005$&$0.396\pm 0.001$&$1125\pm 143$&22.37-26.87\\
\noalign{\smallskip}
 \hline
\end{tabular}
\end{table}
\begin{table}
\caption{A quantitative comparison of galaxy counts in different $R$-surveys}
\label{tab:rdurch}
\begin{tabular}{ccccccc}
\noalign{\smallskip}
\hline
\noalign{\smallskip}
survey&area&slope&amplitude&range\\
\noalign{\smallskip}
\hline
\noalign{\smallskip}
Bertin&$140$&$0.537\pm 0.001$&$579\pm 3$&14.5-19.5\\
NEP$_a$&$1.0$&$0.480\pm 0.020$&$397\pm 30$&14.13-19.38\\
NEP$_b$&$1.0$&$0.498\pm 0.044$&$424\pm 54$&17.88-19.38\\
Jones&$3.0$ &$0.360\pm 0.002$&$448\pm 4$&18.13-22.13\\
NEP$_c$&$1.0$ &$0.368\pm 0.006$&$368\pm 5$&18.88-22.38\\
Metc.$_{91}$&$0.079$&$0.370\pm 0.008$&$432\pm 21$&19.0-23.5\\
Huang&$0.19$&$0.357\pm 0.009$&$397\pm 14$&19.25-22.75\\
Metc.$_{95}$&$0.006$&$0.399\pm 0.017$&$315\pm 55$&21.75-25.25\\
\noalign{\smallskip}
 \hline
\end{tabular}
\end{table}
\section{Galaxy colours}
\label{chap:galcol}
Changing slopes in number counts can lead to identifications of additional
components contributing to the entire sample of objects, but it is
impossible to constrain the nature of such additional components from
number counts alone. Surveys in multiple colours provide further insight.

\subsection{Matching of the $B_j$-, $R$- and $K$-catalogs}
The colours of objects are derived by matching the independently
produced catalogs in $B_j$, $R$ and $K$.
For every entry in the $R$-catalog we searched the $B_j$-catalog
for an object closer than $2''$ to the $R$-position.
If there exists such an entry in $B_j$, The $R$ and the $B_j$ object
are considered to be identical, and an entry in the $B_j-R$-colour catalog
is made. The same procedure was applied to the $K$ and $R$ catalogs
to assemble the $R-K$-colour catalog. Finally, we matched
the $B_j$ catalog with the colour catalog in $R-K$ to get the
full colour information of all objects.
Since the magnitudes derived as outlined in Sect.~\ref{chap:magn} are
``total'' magnitudes, the colours were calculated as the difference
between the magnitudes in each passband.\\
We took $2''$ as the largest distance for the identification of objects
in two catalogs. At distances $>2''$ an increasing number of
object pairs would enter the colour-catalogs which are just a positional
coincidence {\it by chance} of two individual objects in the catalogs of each
passband.
\subsection{The colour-magnitude diagram in $B_j-R$ and $R-K$}
\label{chap:cmd}
\begin{figure}
\resizebox{\hsize}{!}{\includegraphics*[angle=-90]{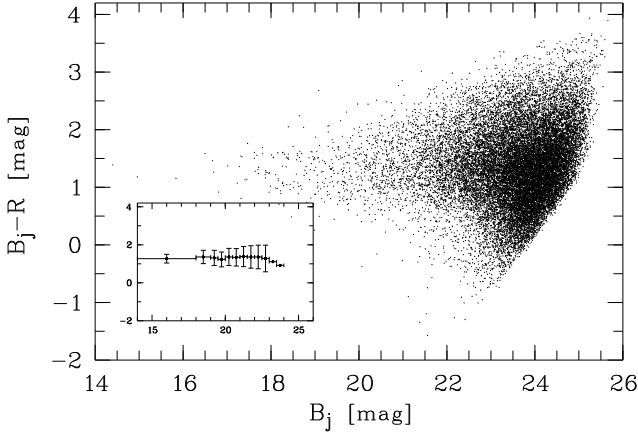}}
\caption{The colour-magnitude diagram of extended sources in $B_j-R$.
In the inset median colours and standard deviations for magnitude
bins indicated by the horizontal bars are plotted.}
\label{fig:fig3}
\end{figure}

\begin{figure}
\resizebox{\hsize}{!}{\includegraphics*[angle=-90]{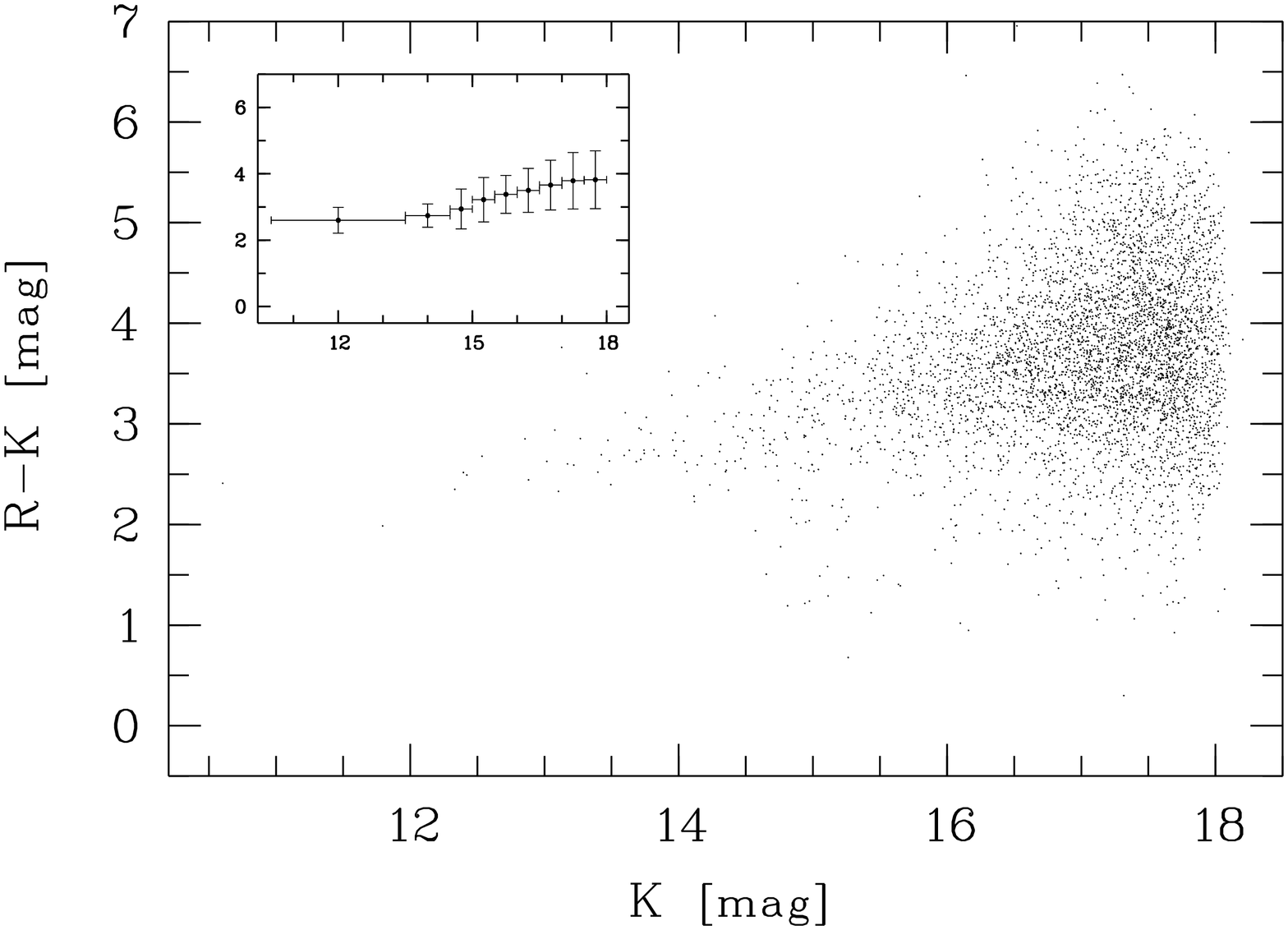}}
\caption{The colour-magnitude diagram of extended sources in $R-K$.
In the inset median colours and standard deviations for magnitude
bins indicated by the horizontal bars are plotted.}
\label{fig:fig4}
\end{figure}

Figs.~\ref{fig:fig3} and \ref{fig:fig4} show the colour-magnitude diagrams of
extended objects in $B_j-R$ and $R-K$, respectively. In those diagrams the
morphological classification is based on the FOCAS-classifier in one filter.
For $B_j-R$ we took the classification in $B_j$, and for $R-K$ the
classification in $R$, since the PSF of the $K$-images is grossly undersampled
(see paper I). Obviously the
statistical extension of the FOCAS-classification for
the number counts (see Sect~\ref{chap:ncounts} and paper I) could {\it not}
be applied  to the individual entries in Figs.~\ref{fig:fig3} and
\ref{fig:fig4}.\\
Because of the misclassification at the faint end $m_{B_j}>23.5\,\mathrm{mag}$
(see Sect.~\ref{chap:class}) some actually extended objects in $B_j-R$ were
marked as point-like sources. Those sources are not represented in
Fig.~\ref{fig:fig3} and the true population is underestimated there.
In $R-K$ this effect starts at $R>22.5\,\mathrm{mag}$ and affects only
the red faint end in Fig.~\ref{fig:fig3} to a rather negligible degree.\\
As already argued in Sect.~\ref{chap:compl} sources with magnitudes down
to $mag_{50}$ in $B_j$, $R$ and $K$ enter the colour-magnitude diagrams.
The sharp cutoffs in Fig.~\ref{fig:fig3} at the right (faint end) and lower
right (blue-faint end) reflect the limited depth of the $B_j$- and
$R$-exposures. Because of the colour-term in the transformation of the
instrumental- to the $B_j$-magnitudes (see Equ.~\ref{equ:cterm}) the cutoff
at the faint end is not parallel to the ordinate.
Due to the large difference in the limiting depth of the $R$- and $K$-survey
the $R$-cutoff in Fig~\ref{fig:fig4} is almost unrecognizable in the sparsely
populated faint red end of the colour-magnitude diagram.\\
Objects with unusual colours were individually inspected for errors in the 
reduction or matching process. Whenever a problem was detected, e.g.\ a 
different splitting of neighboring objects in the two passbands or a problem 
in the determination of the background near bright stars, the objects were 
excluded from the colour-catalogs. In $B_j-R$ and $R-K$ the numbers of removed 
objects is 30 and 20, respectively. Therefore all objects in Figs.\
\ref{fig:fig3} and \ref{fig:fig4}, even if isolated in colour, are {\it
real} objects with good photometry and colours. There are $23\,000$ and
$4\,900$ points in the colour-magnitude diagrams Fig.~\ref{fig:fig3}
and \ref{fig:fig4}, respectively.\\
\subsection{The two-colour diagram}
\begin{figure*}
\resizebox{\hsize}{!}{\includegraphics*[angle=-90]{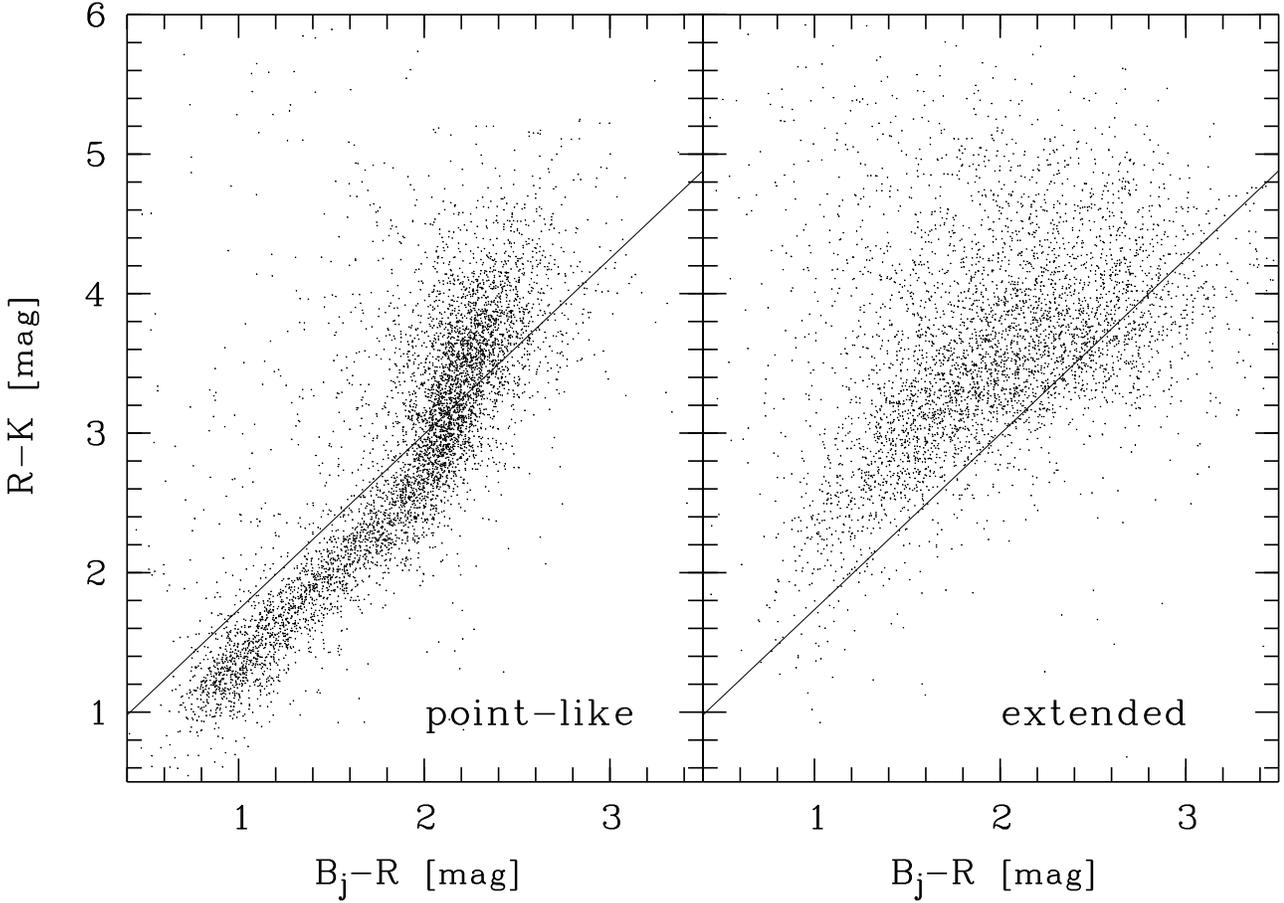}}
\caption{The two colour diagram for extended (right) and point-like (left)
sources The line drawn in both panels divides point-like and extended
objects in the region $B_j-R<1.9$.}
\label{fig:fig11}
\end{figure*}
Fig.\ \ref{fig:fig11} shows the two-colour diagram $B_j-R$ vs. $R-K$
with the point-like and extended objects in the left and right panels,
respectively. As in Figs.\ \ref{fig:fig3} and \ref{fig:fig4} the 
classification into point-like (5\,300 objects) and extended sources 
(4\,200 sources) in Fig.\ \ref{fig:fig11} is based on the 
FOCAS-classification in the $R$-band. As in the colour-magnitude 
diagrams, only objects brighter than $mag_{50}$ in $B_j$, $R$, and $K$ 
were matched and plotted in Fig.\ \ref{fig:fig11}.\\
Most of the point-like objects in the left part of Fig.\ \ref{fig:fig11}
are concentrated along a well defined line of $\sim 0.3\,\mathrm{mag}$
width. While the blue objects are assumed to be halo stars several kpc 
above the galactic disc, the red objects can be identified as faint
{\bf M-dwarfs} within the disc, in immediate vicinity of the sun
(Bahcall \cite{bahcall2}, Robin \& Cr\'{e}z\'{e} \cite{robin},
Baraffe et al.\ \cite{baraffe}). The two populations are not well separated
but are connected with a less densely populated regime at
$B_j-R\sim 1.7\,\mathrm{mag}$.\\
Compared to point-like objects the distribution of extended objects
in the two-colour diagram is much broader. Furthermore, the extended
objects usually have a redder $R-K$ colour. Following Huang et al.\ 
(\cite{huang97}) this allows a separation of point-like and extended 
objects based on colours alone in the blue part to $B_j-R< 1.6\,\mathrm{mag}$ 
at both sides of the line drawn in Fig.\ \ref{fig:fig11}. In the red part 
such a colour based separation is no longer possible, since the locus of 
the point-like objects then lies above the separating line in the colour region
populated by extended objects. Most of the 10 \% contamination of
unresolved galaxies in our list of point-like objects are likely to
have the same distribution in the two-colour diagram as the objects
classified as extended. They would hence be the dominant contribution
in the sparsely populated regime with blue B$_j$-R and red R-K colours.
\subsection{Colour trends in $B_j-R$ and $R-K$}
In order to study the colour evolution we derived the median colour and
its standard deviation in bins of apparent magnitude.
The values are given in Table \ref{tab:BjRcol} and displayed
in the insets of Figs.~\ref{fig:fig3} and \ref{fig:fig4}. The
horizontal bars mark the widths of the bins in magnitude.
For the last two bins in Fig.\ \ref{fig:fig3} a standard deviation could not
be determined, since a significant population of blue objects are beyond
the depth of the $R$-survey. In a similar way the last bin in 
Fig.\ \ref{fig:fig4} is affected by very red objects being missed in the
$R$-survey.
Nevertheless it is possible to put those
$B_j$-objects without $R$-counterparts at the blue end of the colour
distribution to compute the median displayed in Fig.\ \ref{fig:fig3}.\\
The median $B_j-R$  colour of the galaxies remains constant at $\langle B_j-R
\rangle =1.36$
from the brightest objects down to $m_{B_j}=22.3\,\mathrm{mag}$.
Then follows a rapid evolution to bluer colours, reaching $\langle B_j-R
\rangle =0.92$
at the faintest bin. This evolution is triggered by the onset of
the population of {\it faint blue galaxies}, with
$m_{B_j}>22.5\,\mathrm{mag}$ and $\langle B_j-R\rangle <0\,\mathrm{mag}$.
While the full population is present down to $m_{B_j}>23\,\mathrm{mag}$,
only its {\it reddest part} can be seen in Fig.\ \ref{fig:fig3} at
fainter magnitudes, because the bluer ones have no counterparts in the
$R$-survey.\\
In $R-K$ there is a steady trend to redder colours towards fainter magnitudes.
The median of $\langle R-K\rangle =2.60\,\mathrm{mag}$ at the bright end
changes to $\langle R-K\rangle =3.82\,\mathrm{mag}$ at $K=17.8\,\mathrm{mag}$.
In the last two bins the evolution to red colours seems to level off with
the median colour remaining almost constant.
\begin{table}
\caption{The median colours ($med_{B_jR, RK}$) and standard deviations
($\sigma_{B_jR,RK}$) in $B_j-R$ and $R-K$
for magnitudes slices in $B_j$ and $K$, respectively}
\label{tab:BjRcol}
\begin{tabular}{cccccc}
\hline
\noalign{\smallskip}
$B_j$-range &$med_{B_jR}$ & $\sigma_{B_jR}$ & $K_j$-range &
$med_{RK}$ & $\sigma_{RK}$\\ 
\noalign{\smallskip}
\hline
\noalign{\smallskip}
16.0-18.0 & 1.27  & 0.23 & 10.5-13.5 &  2.60 &  0.39 \\
18.0-19.0 & 1.36  & 0.35 & 13.5-14.5 &  2.74 &  0.35 \\
19.0-19.5 & 1.31  & 0.40 & 14.5-15.0 &  2.94 &  0.60 \\
19.5-20.0 & 1.23  & 0.39 & 15.0-15.5 &  3.22 &  0.67 \\
20.0-20.5 & 1.36  & 0.45 & 15.5-16.0 &  3.38 &  0.57 \\
20.5-21.0 & 1.34  & 0.46 & 16.0-16.5 &  3.50 &  0.66 \\
21.0-21.5 & 1.38  & 0.53 & 16.5-17.0 &  3.66 &  0.75 \\
21.5-22.0 & 1.36  & 0.59 & 17.0-17.5 &  3.79 &  0.85 \\
22.0-22.5 & 1.36  & 0.62 & 17.5-18.0 &  3.82 &  0.87 \\
22.5-23.0 & 1.28  & 0.70 & & & \\
23.0-23.5 & 1.11  &      & & & \\
23.5-24.0 & 0.92  &      & & & \\
\noalign{\smallskip}
\hline
\end{tabular}
\end{table}
\subsection{An upper limit to $B_j$ dropouts}
An obvious aim for wide-angle surveys is the derivation of number densities of 
rare classes of objects. Cosmologically important targets are highly 
redshifted targets which can be identified as drop-out objects when the
Lyman edge is redshifted to long wavelengths, out of the bandpass of individual
filters. We determine limits to the surface density of candidates for
high redshifted objects. This allows constraints on the bright end of 
the luminosity function of highly redshifted sources. 
The $K$-band limits are not faint enough to include $R-K$ colour as selection
criterion, and we are confined to the $B_j-R$ index, which does not provide
a unique identification of highly redshifted sources. Nevertheless, it is
interesting for deep wide-angle surveys to determine the surface density of
candidate sources. All $R$-band sources above the completeness limit of our 
sample have a counterpart in the $B_j$ catalog. The only exceptions to this
are a small number of faint sources close to very bright stars, where the
decomposition of faint sources and halo has different efficiency in the two
bands (see section 4.2). There are no true drop-outs in our sample. {\bf The
amount} of the decrement at the Lyman break is discussed controversially.
For non-active  galaxies without Lyman $\alpha$ lines of high EW, a break
of about $2.6\,\mathrm{mag}_{AB}$ magnitudes has been determined
(Steidel et al.~\cite{steidel}). In the filter-system used in our survey,
the Lyman $\alpha$-break is between $B_j$ and $R$ for objects with
$z\sim 3.8$. Thus all objects with $B_j-R>2.6\,\mathrm{mag}$ can be regarded
as candidates for $z\ge 3.8$ galaxies. In Table \ref{tab:highz} we give
the surface density of those candidates in subsurveys with different depth
in $B_j$. The last column of Table \ref{tab:highz} gives the upper limit
in absolute magnitude for an object with $R = R_{lim}$ at $z=3.8$.
The two values refer to the cosmologies $(q_0, \lambda_0) = (0.5,0.0)$ and 
$ (0.1,0.0)$; $H_0 = 50\,\mathrm{km}/\mathrm{sec}/\mathrm{Mpc}$, respectively.
Out of the $9464$ point-like as well as extended
sources down to $R=21.0\,\mathrm{mag}$ we detected no $B_j$-band dropout
over the entire $1\,\mathrm{deg}^2$ field. The reddest objects
are at $B_j-R\sim 3.5\,\mathrm{mag}$. According to Steidel et
al.~(\cite{steidel}) Lyman-break colours $G-R > 3.6\,\mathrm{mag}$
for redshifts $z>4.15-4.45$, depending on the extinction.
Transforming this criterion to our filter set, we expect colours
$B_j-R>3.6\,\mathrm{mag}$ for objects with $z>4.35-4.65$.
Using the formulas given in Steidel et al.(~\cite{steidel}) we can compute from the limit in
apparent $R$-magnitude an upper limit for the absolute magnitude of objects
at $z\sim 4.5$. Depending on cosmology, there are no objects with
$M_{AB}(1280\,\AA) < -25.0, -26.2\,\mathrm{mag}$ for
$(q_0, \lambda_0) = (0.5,0.0), (0.1,0.0); 
H_0 = 50\,\mathrm{km}/\mathrm{sec}/\mathrm{Mpc}$ respectively.
\begin{table}
\caption{The density of high redshift candidates from
the $Ly_{\alpha}$ decrement in the $B_j-R$-colour}
\label{tab:highz}
\begin{tabular}{cccccc}
\hline
\noalign{\smallskip}
$B_{j_{lim}}$ & $R_{lim}$ & area&$N_{cand}$ & 
$\rho _{cand}^{\ast}$& $M_{AB}(1350\AA)$\\
\noalign{\smallskip}
\hline
\noalign{\smallskip}
$<24.375$ &$ <21.775$ &$ 0.3$ &$ 131$ &$ 451$ & -24.00,-25.08\\
$<24.125$ &$ <21.525$ &$ 0.8$ &$ 312$ &$ 371$ & -24.25,-25.33\\
$<23.875$ & $<21.275$ &$ 1.0$ &$ 287$ &$ 287$ & -24.50,-25.58\\
\noalign{\smallskip}
\hline
\end{tabular}
\begin{list}{}{}
\item[$^{\ast}$]The densities are given in $N\mathrm{deg}^{-2}$.
\end{list}
\end{table}
\section{Extremely red objects}
\subsection{The sample of EROs}
Starting from the colour-magnitude diagram Fig.~\ref{fig:fig4} we selected
the population of {\it extremely red objects} or EROs. To give a very
conservative estimate of the surface density of those objects,
we applied additional selection criteria to candidates from
Fig.~\ref{fig:fig4}. With $K < K_{compl}$ we avoid any kind of false or
spurious detection in the $K$-band. Only matches with a separation less than
1 arcsecond from $K$- and $R$-band objects are accepted for EROs (as compared
to $2''$ for Fig.~\ref{fig:fig4}). This criterion rejects matches between
one object from close, but resolved, object pairs in $R$ and their combined,
unresolved counterparts in $K$. Such constellations are caused by the
large pixel size and resolution in $K$ and redden the objects systematically.
Since the merged $K$-object has a different position with respect to
both single objects in $R$, the stronger criterion concerning separation
efficiently removes such mismatches. In $R$ we accepted for the EROs
{\it every} object as counterpart, in contrast to Fig.~\ref{fig:fig4}
where only sources with $mag < mag_{50}$ were considered.
While this might result in matches with non-existing sources in $R$,
{\it no} false EROs are produced since the $R-K$ colour of a solid detection
in $K$ can only become {\it bluer}. Finally, both authors individually
checked the ERO candidates for signs of errors in detection, photometry
and matching of the counterparts in $R$ and $K$. Only objects confirmed
by both authors are considered as EROs.
As $R$-dropouts, objects which only have a lower limit in their $R-K$-colour
we considered only objects with $K < K_{compl}$. To test for errors the
objects were individually checked on the $K$-images.\\
\begin{figure}
\resizebox{\hsize}{!}{\includegraphics*[angle=-90]{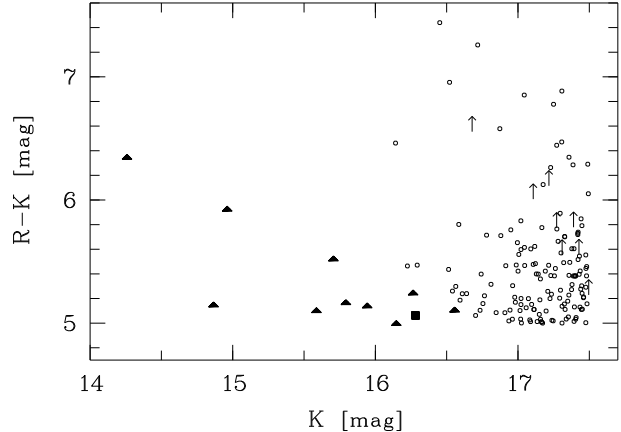}}
\caption{The colour-magnitude diagram of the EROs. Extended and point-like
objects with reliable classification are marked as filled square and triangles,
respectively. $R$-band dropouts are marked with their lower limit.}
\label{fig:fig12}
\end{figure}
\subsection{The surface density of EROs}
Fig.~\ref{fig:fig12} shows the colour-magnitude diagram of the EROs in
our survey. Unfortunately the threshold for EROs in $R-K$  is not
very well defined, and the values vary from $R-K > 5.0$
(Cimatti et al.~\cite{cim}) to $R-K > 6.0$ (Thompson et al.~\cite{thomps}).
Therefore all objects with $R-K > 5.0$ are included in Fig.~\ref{fig:fig12}.
ERO objects without detection in $R$ ($R$-dropouts) are included with their
lower limit in $(R-K)_{lim}$ (computed via $R_{compl}-K$) in
Fig.~\ref{fig:fig12}.
In Table \ref{tab:robj} we give the surface density of the our EROs
for different limits in $K$-magnitude as well as colour $R-K$.
Only EROs in sub-surveys complete down to the $K$-limit specified in
the first column of Table \ref{tab:robj} are taken to compute the surface
densities. The area of those sub-surveys is given in the last column
of Table \ref{tab:robj} (see also paper I).
The $R$-dropouts were taken with their respective $(R-K)_{lim}$ to compute the
surface densities in Table \ref{tab:robj}. The number of $R$-dropouts
is given in Table \ref{tab:robj} within parentheses.\\
Only $11$ EROs out of the $146$ from Fig.~\ref{fig:fig12} are bright
enough to allow a reliable morphological classification in $R$. 
While the only extended object is marked with a filled square, the point-like
sources are given as filled triangles.\\
The two main contributors to our EROs-population are late type stars and
galaxies at high redshift ($z>0.8$). As the available information on
morphology suggests, the bright end is dominated by stellar objects. 
According to Leggett (\cite{leggett}) a colour $R-K>5.0\,\mathrm{mag}$
is expected for stellar types M6 and later. With typical absolute magnitudes
of $M_K=9.5\,\mathrm{mag}$ and $M_K=11.5\,\mathrm{mag}$ for M6-dwarfs and
L-dwarfs, respectively (see Leggett \cite{leggett} and Reid \cite{reid})
and we detect these objects out to a distance of 
$400\,\mathrm{pc}$ and $160\,\mathrm{pc}$.\\
The surface density of extragalactic EROs at the depth of our
survey is completely unknown. Thompson et al.~(\cite{thomps}) give a surface
density of $0.04\,\mathrm{arcmin}^{-2}$ down to $K\le 19.0\,\mathrm{mag}$.
This is more than $5$ times higher than our value at $R-K>6.0$ for
the total population at our limit $K\le 17.5\,\mathrm{mag}$. 
The high surface density of $0.7\,\mathrm{arcmin}^{-2}$ as given by
Eisenhardt et al.~(\cite{eisen}) in their sample down to
$K\le 20.1\,\mathrm{mag}$ gives clues to a fast decline of the density
towards brighter magnitudes. Therefore only a few out of the $16$ objects
in the reddest and deepest interval of Table \ref{tab:robj} might be of
extragalactic origin. Deeper studies of the red population would require data
with both better spatial resolution and wavelength coverage.
\begin{table}
\caption{The density of EROs-object as a function of limits
in $K$-magnitude and $R-K$ colour.}
\label{tab:robj}
\begin{tabular}{llllc}
\hline
\noalign{\smallskip}
 mag & $\rho _{R-K>5.0}^{\ast}$& $\rho _{R-K>5.5}^{\ast}$ &
$\rho _{R-K>6.0}^{\ast}$ & area [deg]$^2$\\
\noalign{\smallskip}
\hline
\noalign{\smallskip}
$K<15.0$ & $0.09$ & $0.06$ & $0.03$&$0.93$\\
$K<16.0$ & $0.21$ & $0.09$ & $0.03$&$0.93$\\
$K<17.0$ & $1.43$\,\,$(1)$ & $0.43$\,\,$(1)$ & $0.21$\,\,$(1)$&$0.91$\\
$K<17.5$ & $5.46$\,\,$(7)$ & $1.91$\,\,$(4)$ & $0.91$\,\,$(3)$&$0.61$\\
\noalign{\smallskip}
\hline
\end{tabular}
\begin{list}{}{}
\item[$^{\ast}$]The densities are given in $10^{-2}\mathrm{arcmin}^{-2}$.
\end{list}
\end{table}
\section{Modeling of the galaxy colour-distributions in $B_j-R$ and $R-K$}
The colour trends seen in the colour magnitude diagrams in
Sect.~\ref{chap:cmd} are caused by a combination of different stellar
populations sampled in galaxies of higher redshift (and therefore observed
at fainter magnitudes) and colour corrections introduced through cosmological
effects, the so called k-corrections. {\bf A method to compute colour distributions
taking into account those effects involves the so called} {\it pure luminosity
evolution} models (Gardner \cite{gardner98}, Pozzetti et al.~\cite{pozzetti}).
The data in our surveys give the unique opportunity to compare two observed
colour distributions with those models. Our large number of sources allows
us to reject models which do not reproduce the observed distribution
with a high significance. Note, that in $R-K$ the comparison is not affected by
an incomplete coverage of red $K$-objects in $R$.
\begin{figure*}
\resizebox{\hsize}{!}{\includegraphics*[angle=-90]{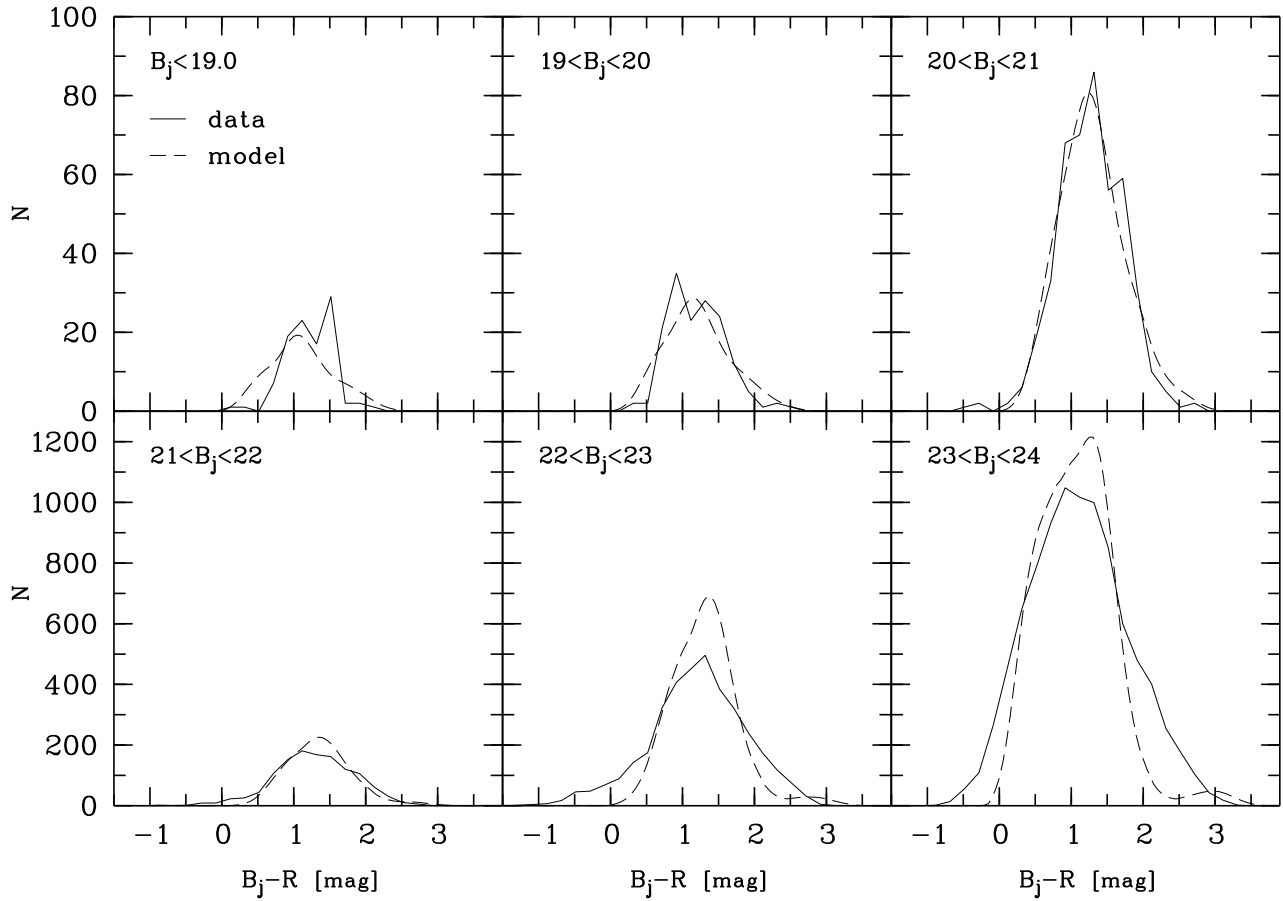}}
\caption{The colour-distribution of galaxies in $B_j-R$
(solid line) in comparison with the distribution of the
model described in the text (dashed line).}
\label{fig:fig9}
\end{figure*}
\subsection{Modeled colour distributions}
\label{chap:subsec}
This modeling is based on theoretical
spectral energy distributions (SEDs) computed with evolutionary synthesis
techniques by Bruzual \& Charlot (\cite{bruzual}).
To derive the colour-distributions from the input {\bf parameters} SEDs,
luminosity function, cosmology and SED-mix (or type mix) we used the
program {\tt ncmod}
developed by Gardner (\cite{gardner98}). The basic characteristics of the
SEDs used in terms of corresponding galaxy type, metallicity,
star formation rate, and epoch of first stars $z_{form}$ are given in 
Tab.\ \ref{tab:sed}. The last column indicates whether passive evolution 
of the galaxy luminosity is taken into account or not. In all models presented
here the cosmological parameters 
$H_0=50\,\mathrm{km}/\mathrm{sec}/\mathrm{Mpc}$,
$q_0=0.02$ and $\lambda_0=0.0$ have been used. Likewise, all models
consider internal absorption by dust according to Wang (\cite{wang}).\\
For the theoretical distribution of $B_j-R$ we assume the $B_j$-based
luminosity function of Loveday et al.\ (\cite{loveday}) and a type
mix of 0\% E1, 10\% E2, 10\% Sa, 15\% Sbc, 45\% Scd, and 20\% Irr.\\
In $R-K$ we show two models using the $K$-luminosity function
from Gardner et al.\ (\cite{gardner97}). The type mix is 16\% E1, 16\% E2,  
28\% Sa, 29\% Sbc, 5\% Scd, and 6\% Irr for model\,1, and 25\% E1, 25\% E2, 
36\% Sa, 10\% Sbc, 3\% Scd, and 1\% Irr for model\,2.
\begin{figure*}
\resizebox{\hsize}{!}{\includegraphics*[angle=-90]{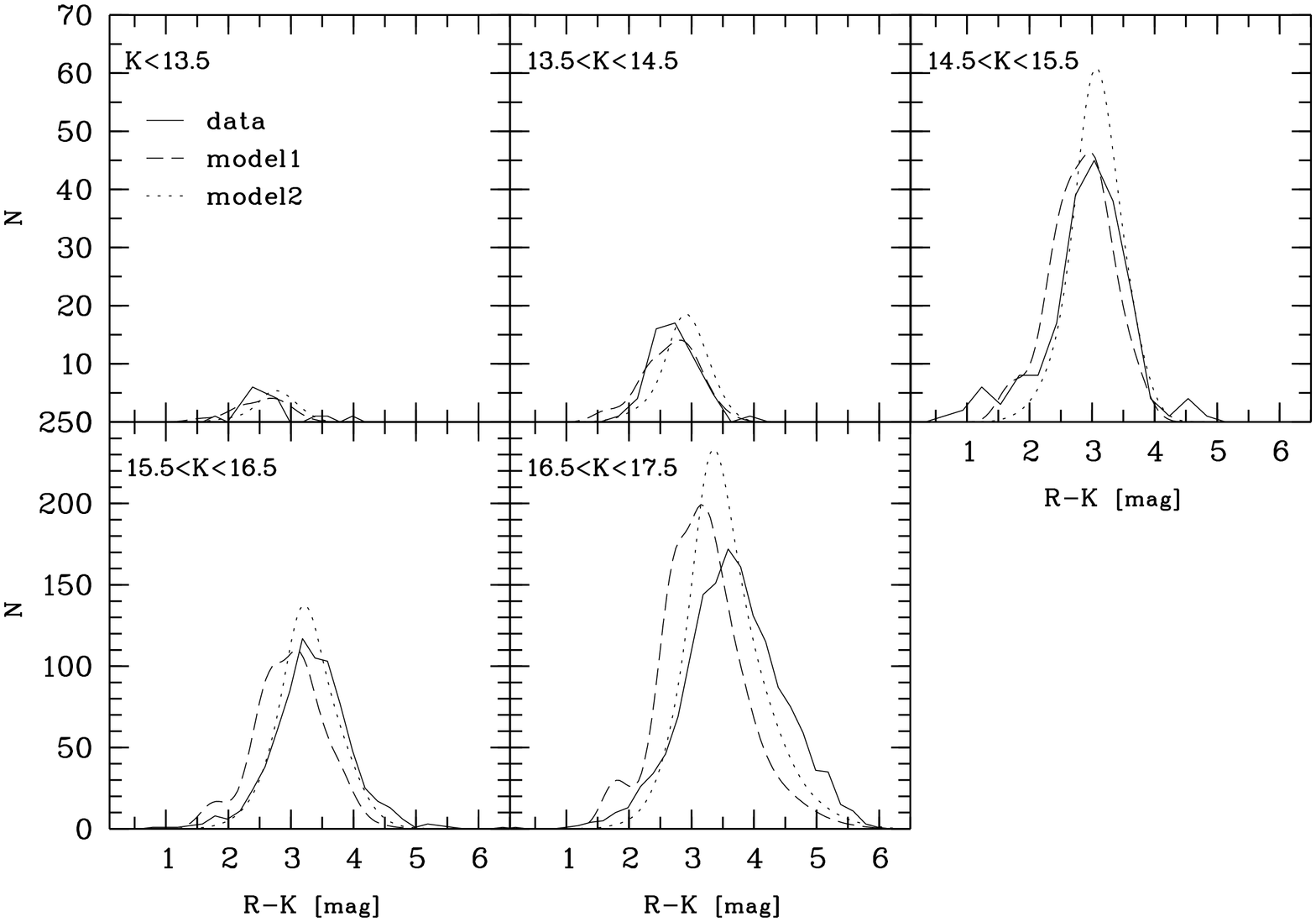}}
\caption{The colour-distribution of galaxies in $R-K$
(solid line) in comparison with the distribution of the
models described in the text (model1: dashed line, model2: dotted line).}
\label{fig:fig10}
\end{figure*}
\begin{table}
\caption{Description of the SEDs used to model the galaxy colours.}
\label{tab:sed}
\begin{tabular}{ccccc}
\hline
\noalign{\smallskip}
type&metalicity&star formation rate&$z_{\rm form}$&evolution\\
\noalign{\smallskip}
\hline
\noalign{\smallskip}
E1&$2.5\times$solar&exp., $\tau=1\,$Gyr& $15$&y\\
E2&$1.0\times$solar&exp., $\tau=1\,$Gyr&$15$&y\\
Sa&$1.0\times$solar&exp.,  $\tau=4\,$Gyr&$15$&y\\
Sbc&$0.4\times$solar&exp., $\tau=7\,$Gyr&$15$&y\\
Scd&$0.2\times$solar&const.&$15$&y\\
Irr&$0.2\times$solar&const.&$1$&n\\
\noalign{\smallskip}
\hline
\end{tabular}
\end{table}
\subsection{Comparison of modeled and observed colour distribution}
Figs.\ \ref{fig:fig9} and \ref{fig:fig10} show the comparison between
the models and the observed colour distribution in $B_j-R$ and $R-K$.
The data are shown in solid, the models in the dashed and dotted lines.
While the observed distributions have been 
corrected for galactic absorption by $\Delta (R-K) = -0.10\,\mathrm{mag}$
and $\Delta (B_j-R) = -0.08\,\mathrm{mag}$ (see Tab.\ \ref{tab:coo}
and Schmidt-Kaler \cite{schmika}), the theoretical
distributions in Figs.\ \ref{fig:fig9} and \ref{fig:fig10} were folded with
a Gaussian of $0.2\,\mathrm{mag}$ FWHM in order to mimic photometric errors.\\
In the comparison between models and data we had to treat the colour
distributions in $B_j-R$ and $R-K$ separately since
none of the models reproduces the colour distribution in $B_j-R$ and $R-K$
simultaneously for a single luminosity function and type mix.\\
Fig.\ \ref{fig:fig9} compares the $B_j-R$-colour distribution of galaxies
in bins of different apparent magnitude $B_j$ to the best fitting
theoretical colour distribution (computed as described in
Sect.~\ref{chap:subsec}). The agreement of the models with the data is
very good throughout the magnitude range covered. There are small
deviations only in the last two panels ($B_j>22\,\mathrm{mag}$).
In the $B_j=22.5\,\mathrm{mag}$ bin an agreement could easily be
reached using a broader Gaussian filter, which is justified by the
larger photometric errors at the faint end. The last panel at
$B_j=23.5\,\mathrm{mag}$ is seriously affected by incompleteness
in both $B_j$ and $R$, which affects the observed colour distribution.\\
Fig.\ \ref{fig:fig10} shows the $R-K$-colour distribution of galaxies
together with the two models outlined in Sect.~\ref{chap:subsec}.
Neither model\ 1 nor model\ 2 is able to reproduce the
observed galaxy colours in $R-K$ over the complete range in apparent
magnitude in $K$. This is caused by the strong evolution of the whole 
population to redder colours as already demonstrated in Sect.\ \ref{chap:cmd}.
While model\ 1 agrees with the observed distribution at the bright
end, but is too blue at the faint end, model\ 2 agrees at the faint end but is
too blue at bright $K$-magnitudes. None of the models can follow the trend of the
observed distributions which become redder by $\sim 1\,\mathrm{mag}$ in the
range $K=13.5-17.5\,\mathrm{mag}$.
\section{Summary}
We have performed a medium deep survey in the filters $B_j$, $R$ and
$K$ with completeness limits $24.25$, $23.0$ and $17.5\,\mathrm{mag}$,
respectively in the central square degree around the Northern Ecliptic
Pole.\\ We have derived the object number counts for point-like and
extended sources in $B_j$ and $R$ and compared our results for the
extended sources to other galaxy number counts found in the literature by
fitting a power law to the data. In both filters we confirm the slope
in dlogN/dm found in other, smaller surveys with values around $0.48$
and $0.37$ in $B_j$ and $R$, respectively.  Differences in the
absolute numbers, represented by the amplitude of the power law fits, 
can largely be attributed to galactic extinction and the limited accuracy of 
reddening corrections.\\
We have determined the colour distribution of galaxies
in $B_j-R$ and $R-K$ in a large range of 10 magnitudes down to
$B_j=24.0$ and $K=18.0\,\mathrm{mag}$, respectively. In $B_j-R$ the
median colour remains constant to $B_j=22.3\,\mathrm{mag}$
and become bluer to fainter levels. This trend to bluer colours marks the 
onset of the so called "faint blue galaxies" (Ellis \cite{ellis}). In
R-K the galaxies become redder from a medium colour $R-K=2.6$ at
$K=12\,\mathrm{mag}$ to $R-K=3.8$ at $K=17.8\,\mathrm{mag}$.\\
For the filter system used in our survey we have given lower limits for the
expected colours (in $B_j-R$) of Lyman-break galaxies at $z\sim 3.8$.
We derive the surface densities for candidates found
in our survey. From the reddest objects in $B_j-R$ found in our survey
we have derived the lower limit for Lyman-break galaxies at $z>4$
to $M_{AB}(12280\,\AA)>-25.0\,\mathrm{mag}$ (for $q_0$, $\lambda_0$, $h_0 =
0.5, 0.0, 0.5$).\\
We have determined the surface density of red objects ($R-K>5.0\,mag$) down to
$K=17.5\,\mathrm{mag}$ on the basis of our large field of view
($0.9\,\mathrm{deg}^2$). Since we are unable to determine the morphology of
most of the objects, the surface densities include late type stars
(M6 and later) as well as the extragalactic
EROs. At the bright end (K$<16.5\,\mathrm{mag}$), which has reliable
morphological classification, point-like objects dominate the sample. 
While the surface density of objects with $R-K>6$ declines by a factor of
twenty from $20.0\,\mathrm{mag}$ to $19.0\,\mathrm{mag}$ (Eisenhardt et 
al.~\cite{eisen}, Thompson et al.~\cite{thomps}, Daddi et al.~\cite{daddi}), 
we find the surface density at $17.5\,\mathrm{mag}$ to be reduced only by an 
additional factor of four.
This comparison shows that either the steep decline in density from
$K=20$ to $19\,\mathrm{mag}$ levels off to $K=17.5$ or that a large 
fraction of our extremely red objects are stars.\\
We have compared the colour distribution of galaxies in $B_j-R$ and $R-K$ with
theoretical colours based on spectral evolution synthesis.
We have shown that it is not possible to find parameters
(e.g. luminosity function and galaxy
type mix) such that both colour distributions are reproduced simultaneously.
While a type mix dominated by late type galaxies reproduces the optical
colour $B_j-R$, the strong trend to redder colours in $R-K$ with increasing
magnitudes can not be reproduced by models. A model which fits the data
at $K=17\,\mathrm{mag}$ is too red at $K=14\,\mathrm{mag}$, a model which fits
at the bright end is too blue at the faint end.
\begin{acknowledgements}
This work was supported by the DFG
(Sonderforschungsbereich 328 and 439) and the
{\it Studienstiftung des deutschen Volkes}.
\end{acknowledgements}
\end{document}